\documentstyle{report}
\input psfig
\psfull

\parskip=\the\medskipamount

\def\etal{{\it et al. \/}}

\def\be{\begin{equation}}
\def\ee{\end{equation}}
\def\bea{\begin{eqnarray}}
\def\eea{\end{eqnarray}}

\renewcommand{\abstract}[1]{{ \footnotesize \noindent {\bf Abstract} #1 \\}}
\renewcommand{\author}[1]{\subsubsection*{#1}}

\begin{document}
\chapter*{Cosmological Constraints on\\
 Terrestrial Planet Formation}

\author{Charles H. Lineweaver\\
School of Physics, University of New South Wales}

\suppressfloats
\section{Cosmoplanetology}

During a marathon, a runner often does not know which position he is in. His nearest competitors are either so far 
ahead or so far behind that he doesn't know where he stands.
We earthlings are in a similar situation. We do not yet have a cosmic context for life.
We don't even know if there are other competitors.
Approximately 5\% of the sun-like stars surveyed 
possess close-orbiting giant planets (Marcy \& Butler 2000) but
we still cannot verify if our Solar System is a typical planetary system.
Cosmology can help.

Since planets form around stars, the planet formation rate in the universe is strongly linked to the star formation rate.
We now have estimates for the star formation rate in the universe (Fig. 1B). 
%
If terrestrial planets (`earths') always formed around stars then the earth formation rate would be
equal to the star formation rate.
But earths are made out of metals and metals are the {\it accumulated} waste product of stars. Thus 
the first stars had no earths and the earth formation rate is correlated with the time integral of the 
star formation rate.
%
The presence of close-orbiting giant planets is both incompatible with the existence
of earths and strongly correlated with high metallicity of the host stars (Fig. 2). 
Thus, there may be a metallicity selection effect:
early in the universe with little metallicity, earths are unable to form for lack of material;
later on, in star forming regions of very high metallicity, giant planets destroy earths.


The central idea of cosmoplanetology is to piece together a consistent scenario based on 
current estimates of the star formation rate of the universe,  the metallicity evolution of the 
star-forming regions of the universe and the most recent observations of extrasolar planets.
The precision of all of these data sets is improving rapidly, but they can already
be combined to yield an estimate of the age distribution of earth-like planets in the universe (Lineweaver 2001).
The earth-like planets in the universe are, on average, 
$1.8 \pm 0.9$  billion years older than the Earth.
If life forms readily on earth-like planets 
-- as suggested by the rapid appearance of life on Earth -- this analysis gives us an age 
distribution for life on such planets and a rare clue about how we 
compare to other life which may inhabit the universe. 


\begin{figure*}[!hp]
\centerline{\psfig{figure=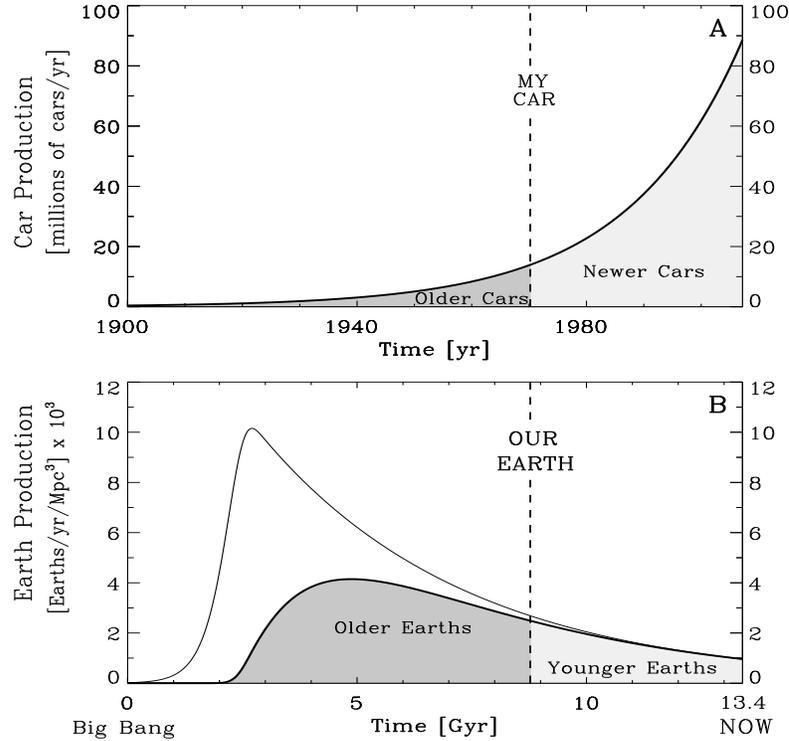,height=10.0cm,width=12.cm}}
\caption{ The World's Car Production vs The Universe's Terrestrial Planet Production.
The top panel (A), shows the world's car production rate -- the number of cars produced per year.
Cars have been produced since about 1900 and the one I drive was built in 1970.
The dark grey area to the left of 1970 is a measure of the number of cars older than mine: only about $20\%$ are older.
I drive an old car. 
The first cars were made about 100 years ago, so the oldest are 70 years older than my car. 
The age of the average car is about 15 years, that is, it was built in about 1985.
The bottom panel (B) shows the universe's terrestrial planet production rate -- the number of earths produced 
per year per cubic megaparsec (See Lineweaver 2001, Fig. 3 for details).
Earths have been produced since about 2.4 billion years after the big bang and our Earth 
was built 4.6 billion years ago, 8.8 billion years after the big bang (Lineweaver 1999).
The dark grey area to the left of $8.8$ billion years is a measure of the number of earth-like planets older than ours,
about $74 \pm 9 \%$ are older. We live on a young planet.
The first earth-like planets were formed about 11 billion years ago so the oldest are about
6.4 billion years older than our Earth.
The age of the average earth in the Universe is $6.4 \pm 0.9$ billion years, that is, it formed about $7$ billion years after 
the big bang. Thus, the average earth in the Universe is $1.8 \pm 0.9$ billion years older than our Earth.
And, if life exists on some of these earths, it will have evolved, on average, $1.8$ billion years
longer than we have on Earth.
For comparison, the thin line is the star formation rate normalized to the earth production rate today.
The time delay between the onset of star formation and the onset of earth production is the 
$\sim 1.5$ billion years that it took for metals to accumulate 
sufficiently to form earths.
}
\label{fig:carproduction}
\end{figure*}


\begin{figure*}[!hp]
\centerline{\psfig{figure=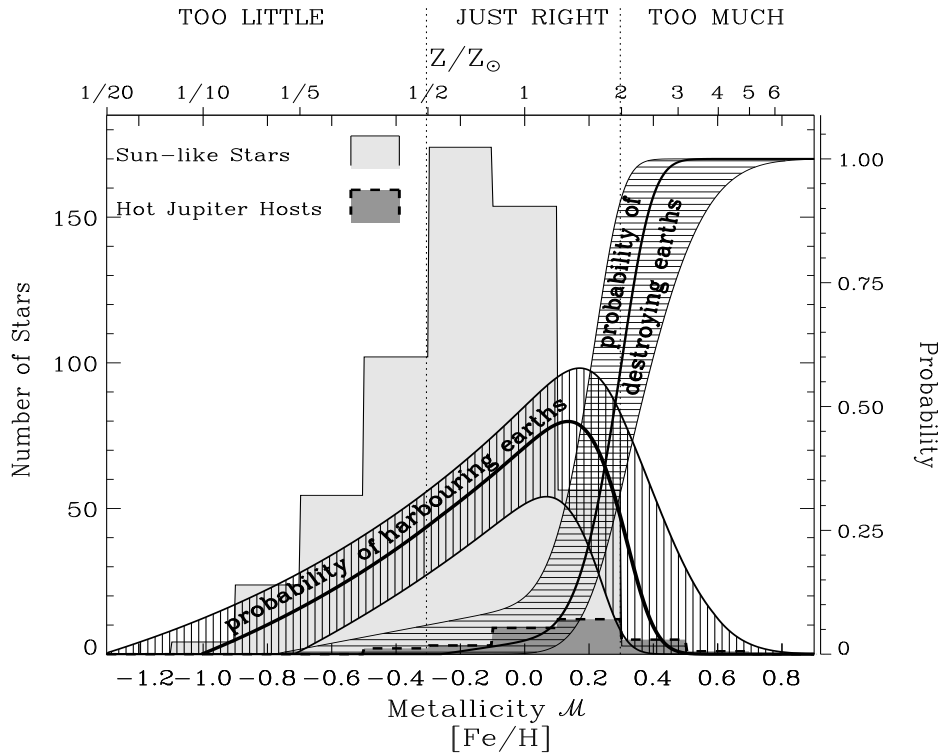,height=9.5cm,width=12.cm,bbllx=30pt,bblly=120pt,bburx=560pt,bbury=610pt}}
\caption{The Metallicity Selection Effect.
If metallicity had no effect on planet formation we would expect the metallicity distribution
of stars hosting hot jupiters (giant, close-orbiting, extrasolar planets, dark grey)
to be an unbiased subsample of the distribution of sun-like stars in the solar neighborhood (light grey). 
However, hot jupiter hosts are more metal-rich.
Hot jupiters have the virtue of being Doppler-detectable but because
they are so massive and so close to the host star and have probably migrated through the habitable zone, they 
destroy or preclude the existence of earths in the same stellar system.
Thus, the probability of destroying earths is the ratio of the dark histogram to the light histogram. 
It is an estimate of the probability that a sun-like star of a given metallicity will have a hot jupiter. 
It is the ratio, as a function of metallicity, of the number of 
hot jupiter hosts to the number of stars surveyed.
The probability of harbouring earths can be constrained by at least three consideration:
1) at high metallicity, earths are destroyed or prevented from forming by the presence of hot jupiters,
2) at zero or very low metallicity, there are not enough metals to form earths,
3) since the Earth and two other earth-like planets (Mars and Venus) exist around the Sun, it is  
reasonable to suppose that terrestrial planets in general have a reasonable chance of forming 
around stars of near solar metallicity.
The probability of harbouring earths shown is
based on these considerations and a production of earths that is
linearly proportional to metallicity.
The upper x-axis shows the linear metal abundance.
The Sun  (${\mathcal{M}}_{\odot} \equiv [Fe/H] \equiv 0$) is more 
metal-rich than $\sim 2/3$ of local sun-like stars and less metal-rich than $\sim 2/3$ of
the stars hosting hot jupiters.
The high value of ${\mathcal{M}}_{\odot}$ (compared to neighboring stars) 
and the low value compared to hot jupiter hosts
is expected if a strong metallicity selection effect exists.
See Lineweaver (2001) Fig. 1 for details.
}
\label{fig:metals}
\end{figure*}

\newpage

\section{Build it and they will come:\\
Earth Production $=$ Life Production?}

The cratering history of the Moon tells us that the Earth underwent an early intense
bombardment by planetesimals and comets from its formation 4.56 Gyr ago until
$\sim 3.9$ Gyr ago.
For the first 0.5 Gyr, the bombardment was so intense (temperatures so high) that
the formation of early life may have been frustrated  (Maher \& Stevensen 1988).
The earliest isotopic evidence for life dates from the end of this heavy bombardment
$\sim 3.9$ billion years ago (Mojzsis \etal 1996). Thus, life on Earth seems to have arisen
as soon as temperatures permitted.

Analogous considerations apply to the origin of life in the universe.
The oldest isotopic evidence for life in the Universe is $ \sim 11$ Gyr.
By `oldest isotopic evidence', I mean, when was the first time in the history of the universe in which the isotopes necessary 
for the formation of an earth-like planet existed in sufficient abundance to form an earth.
The analysis of Lineweaver (2001) indicates that the first earth-like planets probably formed about 2.4 billion 
years after the big bang in the most metal-rich star forming regions of the universe. It took $\sim 1$ billion years 
for the first stars to form and another $\sim 1.4$ billion years for sufficient metallicity to build up.

To interpret the earth production rate of the universe in Fig. 1B as the life production rate 
several assumptions need to be made. Among them are: 1) the dominant harbours for life in the universe are on 
the surfaces of earths in classical habitable zones.
2) life is based on molecular chemistry and cannot be based on just hydrogen and helium.
3) other time-dependent selection effects which promote or hamper the formation of life  
(supernovae rate?, gamma ray bursts?, cluster environments?) are not as important as the metallicity 
selection effect discussed here.

In a lottery the odds of winning can be a thousand to one or a million to one. However, no matter what 
the odds are, the more lottery tickets you buy, the better your chances. Similarly, we do not know how 
likely life is to form on earth-like planets. It may be very likely or it may be next to impossible. 
But whatever the odds, the more earths there are, the more likely life will be to form. Thus, the age 
distribution in Fig. 1B would still represent the age distribtion of life in the 
universe, independent of how likely such events are.

\begin{center}
References
\end{center}

\begin{tabular}{ll}

1. & 
\hbox{\vtop{\raggedright\hangafter=1\hangindent=0mm\hsize 110mm\strut
Lineweaver, C.H. 1999, Science, 284, 1503 
`A Younger Age for the Universe'\strut}} \\ 

2. & 
\hbox{\vtop{\raggedright\hangafter=1\hangindent=0mm\hsize 110mm\strut
Lineweaver, C.H. 2001, Icarus in press, astro-ph/0012399
`An Estimate of the Age Distribution of Terrestrial Planets in the Universe: Quantifying Metallicity as a Selection Effect'}}\\

3. & 
\hbox{\vtop{\raggedright\hangafter=1\hangindent=0mm\hsize 110mm\strut
Maher, K.A. \& Stevenson, D.J. Nature, 331, 612-614 1988
`Impact Frustration of the Origin of Life' \strut}}\\

4. & 
\hbox{\vtop{\raggedright\hangafter=1\hangindent=0mm\hsize 110mm\strut
Marcy, G. W. \& Butler, R.P.  2000, PASP 112, 137-140 
``Millennium Essay: Planets Orbiting Other Suns'' \strut}}\\

5. & 
\hbox{\vtop{\raggedright\hangafter=1\hangindent=0mm\hsize 110mm\strut
Mojzsis, S.J. \etal 1996, Nature, 384, 55-59
`Evidence for life on Earth before 3800 million years ago''\strut}}

\end{tabular}

\end{document}